\documentclass[sigconf,natbib=false,table,hyphens]{acmart}

\setcopyright{none}
\renewcommand\footnotetextcopyrightpermission[1]{}

\usepackage{cite}
\usepackage{amsmath}
\usepackage{graphicx}
\usepackage{algorithm}
\usepackage{algorithmic}
\usepackage{url}
\usepackage{xspace}
\usepackage{booktabs}
\usepackage{multirow}
\usepackage{longtable}
\usepackage{shortcuts}
\usepackage{spverbatim}
\usepackage{subcaption}
\usepackage{float}

\newcommand{\mypar}[1]{\smallskip\noindent\textbf{#1}\xspace}

\begin{document}

\title{\tool: Automated Owner and Abuse Type Tagging for Indicators of Compromise in Threat Reports}

\author{Gibran Gomez}
\affiliation{
  \institution{IMDEA Software Institute}
  \country{Madrid, Spain}
}
\email{gibran.gomez@imdea.org}

\author{Juan Caballero}
\affiliation{
  \institution{IMDEA Software Institute}
  \country{Madrid, Spain}
}
\email{juan.caballero@imdea.org}

\thispagestyle{plain}
\pagestyle{plain}

\begin{abstract}

Cyber Threat Intelligence (CTI) reports often describe 
Indicators of Compromise (IoCs) such as IP addresses, URLs, file hashes, and cryptocurrency wallets involved in cyberattacks.
Those IoCs are typically described in the unstructured report's text, 
or listed at the end of the report with little context, 
limiting their usefulness.
This paper presents \tool, a platform that, given a threat report, 
automatically analyzes its text and 
tags the IoCs it describes with contextual information about 
the threat group and malware family that the IoC belongs to and the 
type of abuse associated with the IoC
(e.g., phishing, sextortion, command-and-control).
\Tool provides a novel LLM-based approach to assign 
owner tags to IoCs using an open-world classification, 
and assigns \numabusetags abuse type tags to IoCs using a 
closed-world classification.
We evaluate \tool on a manually generated 
ground truth of \numgtreports threat reports 
containing \numgtindicators indicators, 
where it achieves an F1 score of 0.94 for owner tagging and 
0.93 for abuse type tagging.
Then, we apply \tool to tag \numwildreports threat reports,
identifying \numiocs IoCs belonging to
\numfamilies malware families, \numgroups threat groups, and 
\numotherowners other entities.
The results show that \tool can tag IoCs
even in reports describing multiple actors and malware families,
enabling the generation of IoC profiles for those entities.

\end{abstract}
\maketitle
\section{Introduction}
\label{sec:intro}

Cyber Threat Intelligence (CTI) reports 
written by security vendors, independent researchers, and incident response teams provide invaluable data about 
attack campaigns, malware families, and threat groups (or threat actors) 
involved in cyberattacks. 
Such threat reports frequently include Indicators of Compromise (IoCs), such as IP addresses, URLs, file hashes, email addresses, and cryptocurrency wallets associated with malicious activity.
However, those IoCs are mentioned as part of the textual description or 
listed at the end of the threat report with little context.
For those IoCs to be actionable, they should be annotated with 
context such as the threat group or malware family they belong to and the 
type of abuse in which they are used 
(e.g., phishing, sextortion, command-and-control).
Such additional context is fundamental for recipients of the IoCs 
in threat sharing platforms~\cite{thomas2016abuse,bouwman2022helping} to understand how to use them 
and also to build threat group profiles
that can be used for threat attribution, threat hunting, 
infrastructure tracking, and 
automated correlation across campaigns~\cite{ethembabaoglu2026apt,saha2026kitten}.

Extracting IoC context from threat reports is a challenging problem.
First, threat reports may include not only IoCs, but also benign indicators 
that should be filtered~\cite{froudakis2025revealing}.
Second, reports often describe multiple threat groups or malware families. 
For example, a report may analyze a malware-based attack where 
a host is infected by a dropper that downloads two other malware samples.
The dropper and the two downloaded samples may belong to 
different malware families 
(e.g., due to pay-per-install schemes~\cite{ppi}).
Thus, we cannot simply manually assign a threat group or malware family to each 
threat report (as done in popular knowledge bases like MITRE ATT\&CK~\cite{attck} and Malpedia~\cite{malpedia}) and assume all IoCs in the report belong to that entity. 
We need to move from a report-level granularity to a per-IoC granularity.
Third, malware families and threat groups are often referred to using different 
names in different reports, and those aliases should be normalized for 
effective threat intelligence sharing.

To address these challenges, we present \tool.
Given a threat report 
(published as a HTML, PDF, Word, or plain text document),
\tool outputs the IoCs in the threat report, 
each annotated with owner and abuse type tags.
The owner tags capture the threat group and malware family that own or control
the IoC, mapping aliases to the same normalized entity tag.
The abuse type tags capture the malicious activities the IoC is used for
(e.g., sextortion, cryptojacking, investment scams). 
In addition, \tool discards indicators in the threat report that are not IoCs
(i.e., benign indicators and false positives from regular expressions)
and classifies hashes (i.e., MD5, SHA1, SHA256)
by the type of artifact they correspond to 
(e.g., file hash or certificate hash).

\Tool implements a novel LLM-based approach to produce 
per-IoC owner and abuse type tags. 
Since new malware families and threat groups continuously appear,
it generates owner tags using an open-world classification 
that creates new tags for previously unknown entities and 
reuses tags for existing ones.
To enable the normalization of aliases to the same entity tag,
it creates an entity database from 
the Malpedia~\cite{malpedia} and MISP~\cite{misp} public knowledge bases.
For generating abuse type tags, 
it implements a closed-world classification 
 using a pre-defined hierarchical taxonomy 
that covers \numabusetags prevalent abuse types and 
is easy to extend.
To improve efficiency and minimize cost, 
\tool also implements a novel generics identification module that 
identifies a large subset of benign indicators without 
requiring an LLM query. 

To evaluate \tool, we generate a ground truth (GT) dataset 
of \numgtreports threat reports, containing \numgtindicators indicators, 
where each IoC is manually annotated with owner and abuse type tags 
and benign and false indicators are labeled as such. 
We evaluate \tool on the GT using a variety of open models run locally.
Using gpt-oss-20b, \tool achieves an F1 score of 0.94 for owner tagging and 
0.93 for abuse type tagging, 
highlighting the accuracy of the proposed approach.
\Tool is model-independent.
Configuring it to use commercial frontier models may further improve accuracy,
but would make our evaluation less replicable and more costly. 

We apply \tool to \numwildreports threat reports
containing \numcandidates candidate indicators, 
collected from three public datasets: 
Malpedia~\cite{malpedia}, AnnoCTR~\cite{annoctr}, and PRISM~\cite{lance}.
\Tool tags \numiocs (77.0\%) candidate indicators as IoCs, 
4,117 (20.3\%) as benign, and 
563 (2.8\%) as invalid indicators. 
The most commonly assigned abuse type tags are
\ttag{command\_and\_control} assigned to 26.9\% IoCs, 
\ttag{malware\_distribution} (22.0\%), and 
phishing (12.6\%).
Among the \numiocs IoCs, \tool assigns an owner tag to \numiocsowner (70.3\%),
with the rest belonging to abuse types where a threat group or malware family 
is rarely assigned, such as extortions, phishing, and other scams. 
The \numiocsowner IoCs with owner tags belong to \numentities entities:
\numgroups threat groups, \numfamilies malware families, and 
\numotherowners other entities.
Of those, 473 (51.8\%) entities are not part of the original entity database.
\Tool allows creating threat group and malware family profiles 
that capture IoCs from the entity appearing in different reports. 
For example, it identifies 369 IoCs belonging to APT28 from 14 threat reports
and 189 IoCs related to Emotet from 8 reports.

Our contributions are summarized as follows:

\begin{itemize}

\item We present \tool, a CTI platform that, given a threat report, 
automatically tags IoCs in the report with owner and abuse types.
\Tool provides 
an open-world LLM-based classifier for assigning owner tags, and
a closed-world LLM-based classifier for assigning \numabusetags abuse type tags.

\item We produce a ground truth of \numgtreports threat reports, 
containing \numgtindicators indicators, 
with each indicator labeled as an IoC or not, and 
with IoCs being annotated with owner and abuse type tags extracted 
from the report's text.

\item We develop a generics filtering module to remove 
a large fraction of benign indicators identified by regular expression 
extraction tools efficiently, without querying the LLM.

\item We evaluate \tool on the GT, showing that it achieves an 
F1 score of 0.94 for owner tagging and 0.93 for abuse type tagging.
We apply \tool to \numwildreports reports,
where it assigns owner tags to \numiocs IoCs belonging to \numentities entities.

\end{itemize}
\section{Problem Definition}
\label{sec:problemdef}

The goal of this work is, given an unstructured threat report 
written in natural language
(e.g., published as an HTML webpage, a PDF or Word document, or as plain text)
to generate a structured report with the indicators of compromise (IoCs)
present in the document (e.g., domains, URLs, cryptocurrency addresses),
each annotated with owner and abuse type tags
that provide context on the IoC.

In this work, an \emph{indicator} is a pair with an indicator type and an indicator value 
such as (fqdn, malicious.com).
We group indicator types into four categories:
networking (e.g., IP addresses, domains, URLs),
contact (i.e.,  emails, TOX identifiers),
hashes of artifacts (e.g., files, certificates), and
addresses of different blockchains (e.g., Bitcoin, Ethereum).
Indicators can be benign or malicious. 
For example, a domain referenced in a threat report may belong to the victim 
or to the attacker.
We use \emph{indicators of compromise} (IoCs) to refer exclusively to malicious indicators,

\subsection{Ownership}
\label{sec:ownership}

Ownership of an indicator can be an elusive concept, as indicators may have multiple owners. 
For example, an IP address could be considered to be owned by the 
Regional Internet Registry (RIR) that has been delegated by IANA to handle the block of 
addresses the IP belongs to; 
by an ISP or cloud hosting provider to whom the RIR leases a sub-block of addresses 
containing the IP address; or 
by a client of those services that leases the IP address to connect to the 
Internet or to host a cloud VM.

We consider ownership at the finest granularity, 
i.e., the owner is the entity at the end of the leasing chain,
who controls how the indicator is (ab)used.
For example, for IP addresses, the owner is the client of an ISP or 
cloud hosting provider that is responsible for how the IP address is used.

Our focus is on tagging IoCs (i.e., malicious indicators). 
We define the owner of IoCs as the threat group or malware family that uses 
the indicator for its malicious activities.
Since threat groups may use multiple malware families, 
\tool may output both threat group and malware family owner tags 
for the same IoC. 
For example, domain malicious.com may be assigned the tags
\ttag{command\_and\_control}, \ttag{plugx}, and \ttag{apt8} 
to indicate that it is a C2 domain used for the PlugX malware by the 
threat group APT8.

Some malicious activities such as scams and extortions are rarely associated with a threat group or 
a malware family.
Instead, they are typically clustered into campaigns with unknown attribution. 
\Tool does not attempt to identify an owner for scams and extortions.
It annotates those IoCs with an abuse type specifying the type of 
scam (e.g., \ttag{investment}, \ttag{techsupport}) or 
extortion (e.g., \ttag{sextortion}, \ttag{ddos}).
Next, we detail ownership for different indicator types in the context of this work.

\mypar{Domains.}
The owner of a domain is generally the entity that has registered the apex from a domain registrar.
However, some services (e.g., blogs or webpage builders) lease subdomains under their apex to third parties~\cite{desilva2021compromised,ahmed2026blockmenot}. 
For \emph{subdomain-leasing apexes}, the owner of the apex is a service,
but the owner of the subdomain is a user of the service.
For example, \textit{blogspot.com} belongs to Google's Blogger service and 
should be considered benign by \tool.
But, its subdomain \textit{malicious.blogspot.com} is leased to a user of 
Blogger and could belong to a malicious entity. 

\mypar{URLs.}
A URL typically belongs to the owner of the domain that hosts it. 
However, some services host user-uploaded content under their apexes~\cite{desilva2021compromised,ahmed2026blockmenot}. 
For \emph{URL-leasing apexes}, the owner of the URL is the user of the service. 
For example, \textit{https://github.com/} belongs to GitHub and should be considered benign by \tool.
But, \textit{https://github.com/SoomeUser/SomeRepo/file.exe} points to 
content uploaded by a user of GitHub and could belong to a malicious entity. 

\mypar{IP addresses.}
The owner of an IP address is the client of an ISP or 
cloud hosting provider that uses the IP address to access the Internet or host some content.

\mypar{Emails.}
An email address belongs to the owner of its domain, except if the domain belongs to an 
email service provider (e.g., \textit{gmail.com}, \textit{protonmail.com}) in which case the 
owner is the user of the email service.

\mypar{Blockchain addresses.}
Blockchain addresses can be generated by individuals,
but also by services that lease them to their users. 
For example, a cryptocurrency exchange may provide each user with a service-controlled wallet so that 
the user can receive coins to be cashed out at the exchange.
For service-provided addresses, we consider the owner to be the user to which the service assigns 
the address.
For example, if a ransom note contains a Bitcoin payment address in an exchange, we consider the 
owner to be the ransomware family, not the exchange.

\mypar{Hashes.}
For artifact hashes (e.g., files, certificates), 
the owner is the threat group or malware family to which the artifact belongs.

\subsection{System Properties}
\label{sec:properties}

\tool provides five properties.
Given a threat report, \tool first identifies candidate indicators in the report using 
a regular-expression-based tool~\cite{iocparser,iocextract,iocsearcher}.
Three refinement properties allow to 
(R1) discard candidate indicators that are not real IoCs,
but rather false positives (FPs) due to errors in the regular expressions; 
(R2) determine whether the indicator is an IoC or benign; and
(R3) classify hashes (i.e., \ioc{md5}, \ioc{sha1}, \ioc{sha256}) 
by the type of artifact they correspond to 
(e.g., \ioc{file.md5}, \ioc{certificate.sha1}).

The core of our approach assigns tags to IoCs capturing:
(C1) \textbf{Who} owns or controls the indicator 
(i.e., malware family, threat group), and
(C2) \textbf{What} type of abuse the indicator is used for 
(e.g., ransom, scam, extortion).
Capturing the owner of an indicator (C1) requires an open-world classification 
since new malware families and threat groups continuously appear. 
It is unfeasible to list all possible entities exhaustively and to keep 
any such list up-to-date.
In contrast,
we use a closed-world classification
and a pre-defined hierarchical taxonomy
to determine the abuse type (C2).

\begin{table}[t]
\centering
\caption{For different datasets, number of reports mentioning multiple threat groups or malware families.}
\label{tab:annoctr}
\resizebox{\columnwidth}{!}{
\begin{tabular}{llrrrr}
\toprule
\textbf{Dataset} & \textbf{Entities} & \textbf{Reports} & \textbf{Multiple} & \textbf{One} & \textbf{Max} \\
\midrule
misp-malpedia & Families & 17,489 & 3,223 (18.4\%) & 14,266 (81.6\%) & 51 \\ misp-threat-actor & Groups & 2,857 & 136 ( 4.8\%) & 2,721 (95.2\%) & 24 \\ AnnoCTR & Group\&Fam & 120 & 83 (69.2\%) & 16 (13.3\%) & 18 \\ \bottomrule
\end{tabular}
}
\end{table}

\subsection{Per-IoC Granularity}
\label{sec:granularity}

Datasets like Malpedia~\cite{malpedia},
MISP Threat Actors~\cite{misp-threat-actor}, 
and AnnoCTR~\cite{annoctr}
manually tag threat reports with the malware families and threat groups mentioned in the report.
However, as illustrated in Table~\ref{tab:annoctr}, it is common that 
the same report may mention multiple malware families and threat groups.
For example, of the 17,489 malware-related threat reports in Malpedia,
3,223 (18.4\%) mention multiple malware families,
with some reports mentioning up to 51 malware families. 
In AnnoCTR, the percentage of reports mentioning multiple  
malware families or threat groups reaches 69.2\%, since this dataset contains many reports 
summarizing the threat ecosystem (e.g., quarterly reports by security vendors). 

Report tags cannot be assigned to all IoCs mentioned in the report because different IoCs in the same report may belong to different malware families or threat groups.
Furthermore, threat reports frequently mention benign indicators that are not part of the infrastructure of threat groups or malware families.

\begin{figure}
\centering
\includegraphics[width=\linewidth]{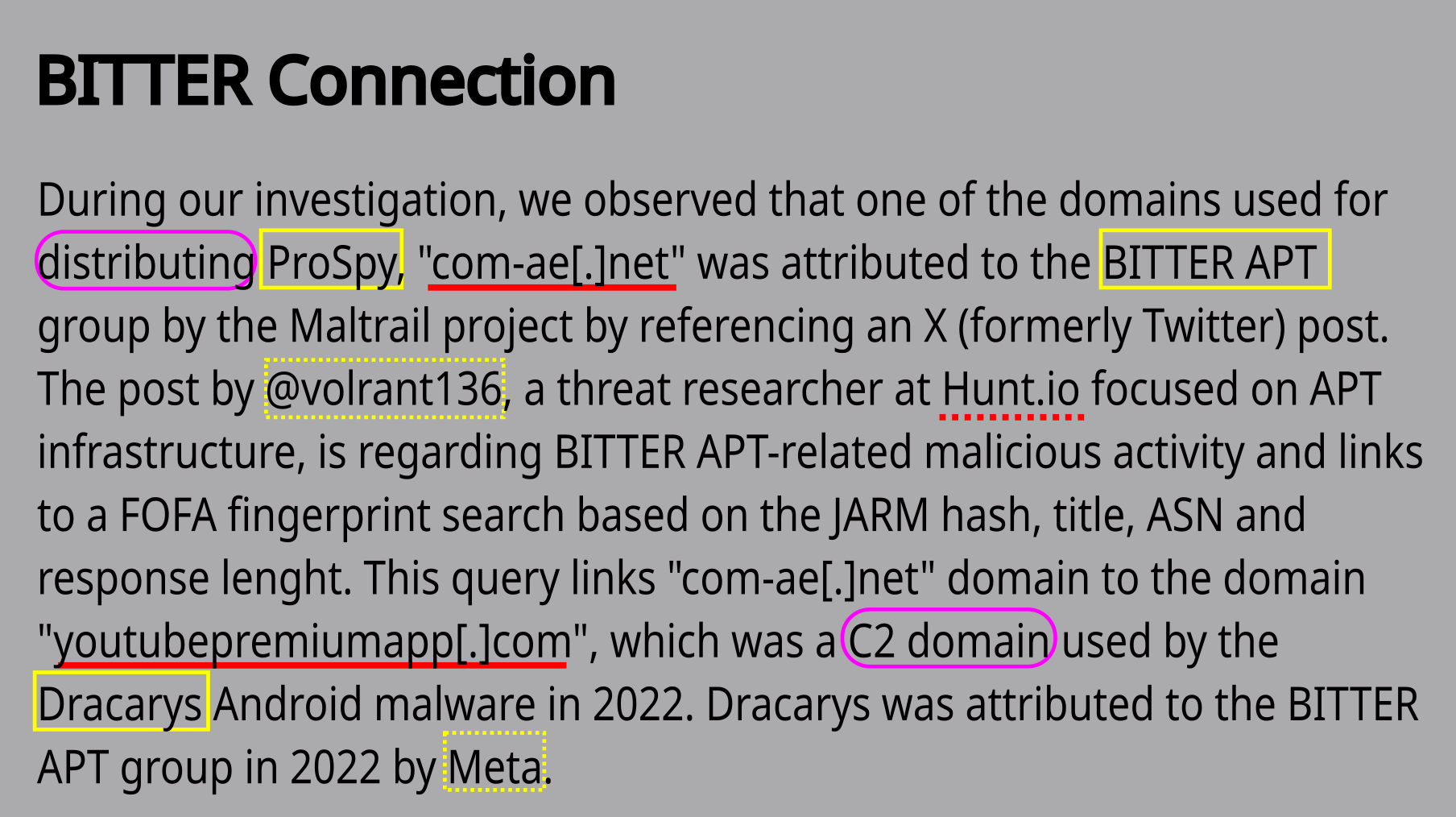}
\caption{
Paragraph of a threat report analyzing two IoCs (underlined with solid red lines)
related to two malware families (\ttag{prospy}, \ttag{dracarys}) and 
a threat group (\ttag{bitter}) (in solid yellow squares).
The paragraph also mentions a benign indicator (dashed red line)
related to a benign actor (dashed yellow square),
and the abuse type of the second IoC (in a rounded purple square).
}
\label{fig:running_example}
\end{figure}

\Tool automates the tagging process at the IoC granularity, 
avoiding errors introduced when applying report tags to all indicators they contain.
Take the paragraph in Figure 1 as an example, 
extracted from a threat report~\cite{lookout}.
It mentions two IoCs 
(\url{com-ae[.]net}, \url{youtubepremiumapp[.]com}) and
one benign indicator (\url{Hunt.io}).
The two malicious domains are linked to the BITTER APT threat group, 
but each is associated with a different malware family
(ProSpy, Dracarys).
The paragraph also mentions some benign entities 
such as \emph{volrant136} and \emph{Meta}.
\Tool will output three indicators.
It will produce three tags for IoC \url{com-ae.net}:
\ttag{malware\_distribution} to capture its abuse type and 
\ttag{prospy} and \ttag{bitter} to capture ownership by a
malware family and a threat group, respectively.
Similarly, it will produce three tags for IoC \url{youtubepremiumapp.com}:
\ttag{command\_and\_control}, \ttag{dracarys}, and \ttag{bitter}.
Finally, it will tag \url{Hunt.io} with \ttag{notanioc} to 
indicate that it is a benign indicator.

\begin{figure}[t]
\centering
\includegraphics[width=\linewidth]{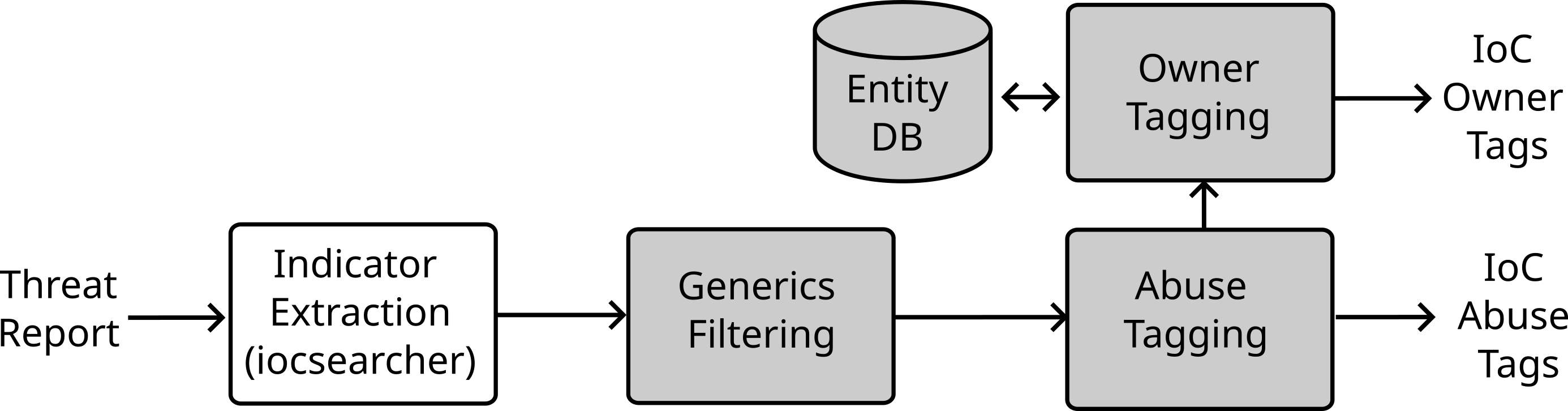}
\caption{\tool architecture. Gray modules were developed for this work.}
\label{fig:arch}
\end{figure}

\section{Approach Overview}
\label{sec:arch}

Figure~\ref{fig:arch} summarizes \tool's architecture.
Given a threat report, 
it annotates the IoCs in the report with abuse type and owner tags.
\Tool comprises four modules: 
\emph{Indicator Extraction}, 
\emph{Generics Filtering}, 
\emph{Abuse Tagging}, and 
\emph{Owner Tagging}.

The indicator extraction module extracts the plain text from the threat report 
file, which may be an HTML webpage or a PDF or Word document, 
and identifies indicators in the text 
using the \iocsearcher indicator extraction tool~\cite{iocsearcher}.
For each indicator found, \iocsearcher returns
the start offset,
the indicator value as it appears in the text (e.g., test(AT)example(DOT)com),
and the defanged indicator value (e.g., test@example.com).
\iocsearcher has been shown to have a very high recall
(i.e., low false negatives) but lower precision (i.e., false positives) 
since regexps can occasionally wrongly identify indicators 
(e.g., identify \emph{file.py} as a domain name rather than a filename) and 
they cannot determine whether an indicator is benign or 
malicious~\cite{goodfatr,froudakis2025revealing}. 

The indicators extracted by \iocsearcher go through the 
generics filtering module, described in Section~\ref{sec:generics},
which aims to filter a subset of the benign indicators 
without analyzing the text. 
Any benign indicator it identifies does not need to be tagged,
thus saving LLM queries and improving tagging process efficiency.

The abuse tagging module,
detailed in Section~\ref{sec:abuse},
takes the filtered indicators and  tags them with abuse type tags.
This module implements a closed-world LLM-based classifier 
that selects the best tags
from predefined taxonomies, 
detailed in Section~\ref{sec:gt}.
It first classifies each indicator into one of three classes:  
a false positive of \iocsearcher, 
a benign indicator (missed by the generics filtering module), or 
an IoC.
This step addresses requirements R1 (i.e., filtering invalid IoCs) and 
R2 (i.e., separating IoCs from benign indicators).
Then, it tags IoCs with abuse types, satisfying requirement C2. 
For hashes, it also determines the type of IoC the hash corresponds to,
satisfying requirement R3.

The owner tagging module produces a set of owner tags capturing the entities
that own the IoC,
divided into three groups: threat groups, malware families, and others
(e.g., individuals, countries),
satisfying requirement C1.
This module comprises two steps detailed in Section~\ref{sec:owner}.
First, it queries the LLM to propose one or more owner names in an open-world fashion, 
i.e., without constraining the names the model may output. 
If the LLM outputs an owner name,
an additional refinement step normalizes aliases for the same entity.
For this, the module searches for similar entity names
in a pre-generated entity database (DB), 
described in Section~\ref{sec:entities},
which includes popular malware families and threat groups, and their aliases.
Second, it queries the LLM to check if the proposed name corresponds to any 
of the candidate entities. 
If so, the known entity tag is assigned. 
Otherwise, it creates a tag for the proposed name and 
a new entity is added to the \db.

\section{Datasets \& Tags}
\label{sec:datasets}

\begin{table}[t]
\centering
\caption{Sources used to build the entity database.}
\label{tab:entities}
\begin{tabular}{lrrr}
\toprule
\textbf{Source} & \textbf{Malware Families} & \textbf{Threat Groups} \\
\midrule
misp-malpedia~\cite{misp-malpedia} & 3,683 & - \\
misp-ransomware~\cite{misp-ransomware} & 2,095 & - \\
misp-threat-actor~\cite{misp-threat-actor} & - & 979 \\
\hline
All & 5,778 & 979 \\
\bottomrule
\end{tabular}
\end{table}

\subsection{Entity Database}
\label{sec:entities}

A challenge when assigning owner tags to IoCs is that an entity 
may be referred to using different aliases.
Some aliases are syntactically similar
(e.g., \emph{Lazarus} and \emph{Lazarus Group}),
but others are not 
(e.g., \emph{Killdisk}, \emph{Sandworm}).

To avoid creating different tags for the same entity,
we create an \db containing popular malware families and threat groups, 
and their aliases.
Each entry in the \db is an entity object containing 
a unique canonical name, a set of aliases, and
optional information about the entity such as 
its type (e.g., malware family or threat group), 
a textual description of the entity, or 
the type of abuse in which the entity participates.
The canonical name is a textual unique identifier (e.g., \ttag{lazarus}) 
using a limited set of characters 
(i.e., lowercase, underscore, period).
The canonical name is used as the entity's tag. 

To build the \db, we leverage the sources in 
Table~\ref{tab:entities}.
We obtain malware families and threat groups from
the MISP open source threat intelligence platform~\cite{misp}.
MISP contains lists  of entities (called \emph{clusters})
contributed by different organizations. 
We use three frequently updated MISP clusters: 
\emph{misp-malpedia}, maintained by the Malpedia~\cite{malpedia}
administrators with the malware families in that knowledge base; and 
\emph{misp-ransomware} and \emph{misp-threat-actor}, 
both maintained by the MISP administrators
with popular ransomware and threat groups, respectively.
Entities in these clusters include aliases. 
Malpedia does not have its own threat group 
knowledge base, instead using \emph{misp-threat-actor}.

\mypar{Search functions.}
The \db provides three search functions that, given a name string,
check if an entity with that canonical name, alias, or other similar 
names exists in the DB.
The \emph{search\_entity} function 
returns the entity with the given canonical name, or 
NULL if the input is not a canonical name. 
The \emph{search\_alias} function checks if the given string is
a known alias for any entity in the DB. 
The same alias may be used to refer to multiple 
entities in MISP and Malpedia. 
This function does not try to resolve such conflicts. 
Instead, it returns a list with all entities in the DB with 
that alias, or an empty list if the alias is unknown.
Finally, the \emph{search\_k\_similar} function performs a k-nearest-neighbor 
search that returns the $k$ entities with a name (i.e., canonical or alias) 
most similar to the input string.
This search computes a similarity score between the input string and each 
name in the \db using the
{S\o rensen}-Dice coefficient for bigrams~\cite{sorensen,dice} 
and returns the entities for the top $k$ most similar names. 
To speed up the search, the bigrams of the canonical labels and aliases
in the \db are precomputed and indexed.
The search first selects the names in the \db
sharing at least one bigram with the input string.
The similarity score is only computed between the input string and this
subset of candidate names, drastically reducing the search time. 
The candidates are sorted by decreasing similarity
and the top $k$ candidates are selected.
The set of entity objects corresponding to the selected candidates is returned.
Less than $k$ entities can be returned
if there were fewer than $k$ names in the \db that shared 
at least one bigram with the proposed label, 
or if multiple selected candidate names were aliases for the same entity.
   
\mypar{Incompleteness.}
No \db can be complete, 
as new malware families and threat groups keep appearing. 
When \tool identifies a previously unknown entity, 
that entity is added to the \db and can be contributed 
back to the original MISP clusters. 

\begin{figure}[t]
    \centering

        \begin{minipage}[t]{0.48\linewidth}

        \vspace{0pt}

        \centering
        \begin{subfigure}[t]{\linewidth}
            \includegraphics[width=\linewidth]{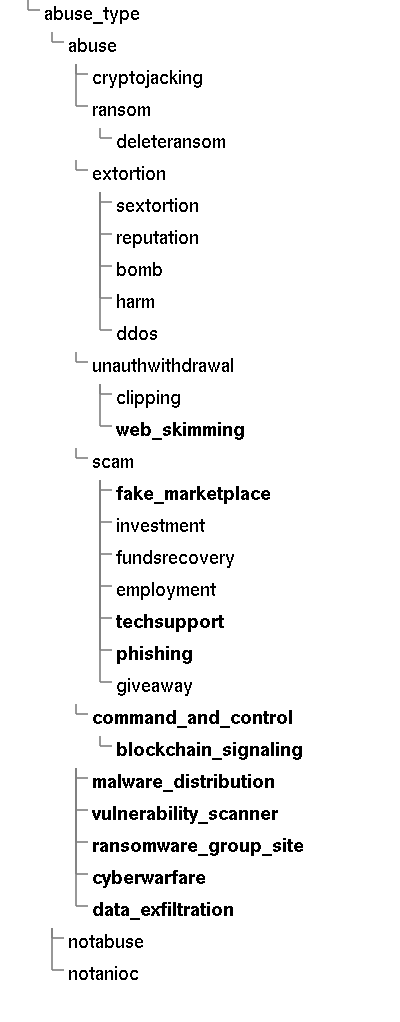}
            \caption{Abuse types}
            \label{fig:abuseTags}
        \end{subfigure}

        \vspace{0.5cm}

        \begin{subfigure}[t]{\linewidth}
            \includegraphics[width=\linewidth]{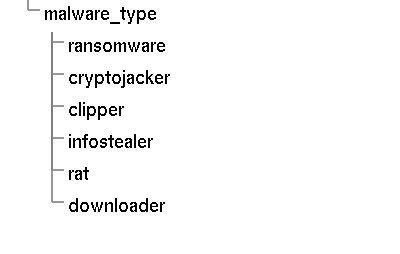}
            \caption{Malware types}
            \label{fig:malwareTags}
        \end{subfigure}

    \end{minipage}
    \hfill
        \begin{minipage}[t]{0.48\linewidth}

        \vspace{0pt}

        \centering
        \begin{subfigure}[t]{\linewidth}
            \includegraphics[width=\linewidth]{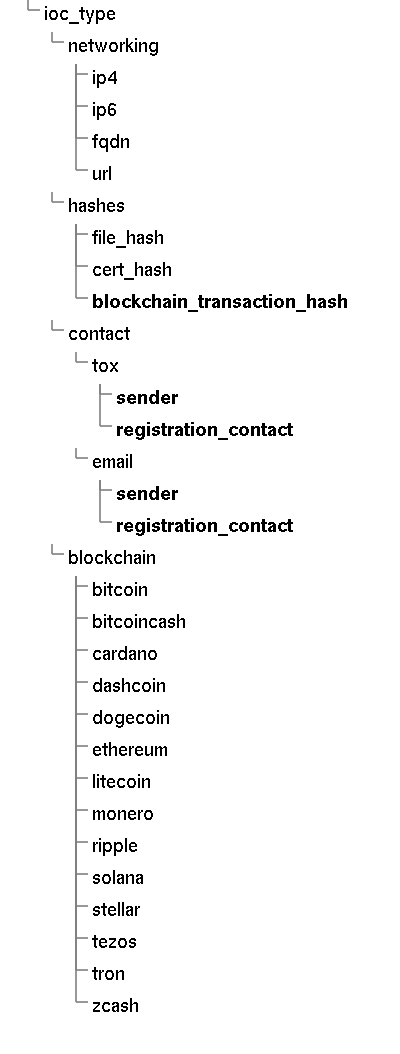}
            \caption{IoC types}
            \label{fig:ioctypeTags}
        \end{subfigure}

        \vspace{0.5cm}

        \begin{subfigure}[t]{\linewidth}
            \includegraphics[width=\linewidth]{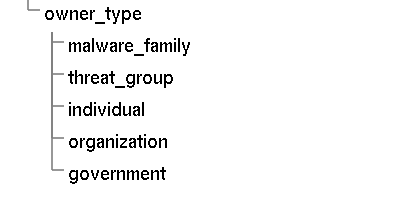}
            \caption{Owner types}
            \label{fig:ownerTags}
        \end{subfigure}

    \end{minipage}

    \caption{Tags used. Bolded tags are those introduced during GT generation.}
    \label{fig:tags}
\end{figure}

\subsection{Ground Truth \& Tags}
\label{sec:gt}

To evaluate \tool, we need a ground truth (GT) dataset of threat reports where each IoC in the reports has been assigned abuse, owner, and indicator type tags. 
No such GT exists, so we created it. 

To generate the GT, we first collected 96 threat reports with analyses of malware and cybercrime operations from a variety of sources including websites of security vendors (e.g., ESET~\cite{eset-welivesecurity}, Cisco~\cite{cisco-talos}), blogs of independent researchers (e.g., Crypto-Ransomware Digest~\cite{id-ransomware}), and cybersecurity news outlets (e.g., BleepingComputer~\cite{bleepingcomputer}).
We also included 4 reports from blockchain-related blogs unrelated to cybercrime.
These reports only contain benign indicators (i.e., no IoCs) and are useful to evaluate potential \tool FPs.
The \numgtreports reports span a period of ten years (2016–2026) and 
contain \numgtindicators candidate indicators identified by \iocsearcher.

Then, we created initial taxonomies of abuse types and indicator types. 
For the abuse types, we started with the taxonomy proposed by 
Gomez et al.~\cite{cleanupthemess}. 
For IoC types, we started with those supported by \iocsearcher, 
with additional types for common artifacts identified by hashes 
(e.g., files, certificates, blockchain transactions).
The taxonomies are illustrated in Figures~\ref{fig:abuseTags} and~\ref{fig:ioctypeTags}, where tags in the initial taxonomy are those that are not bolded.
The taxonomies are hierarchical with deeper levels capturing more specific types.
This design accommodates possible incompleteness and facilitates extensions.
This is particularly important for abuse types, 
since any abuse taxonomy is necessarily incomplete as cybercriminals 
frequently introduce new types of abuses. 
For example, if a threat report refers to a type of scam that is not captured 
by any of the 7 \ttag{scam} subtypes in Figure~\ref{fig:abuseTags}, 
then the parent \ttag{scam} tag is used to indicate that it is another type 
of scam.

Two analysts reviewed the \numgtreports reports, tagging each indicator. 
The analysts were told to label based exclusively on the text. 
They should first classify the indicators into FPs of \iocsearcher (\ttag{notanioc}), benign indicators (\ttag{notabuse}), and IoCs (\ttag{abuse}).
Then, they should label IoCs using the tags in the initial abuse type and indicator type taxonomies, as well as in the \db.
The analysts were allowed to propose new tags if the existing ones 
were not sufficient (e.g., for new abuse types and owners). 
After labeling the reports, the analysts discussed disagreements until they reached a consensus, and normalized the names of the new tags. 

During the GT construction, 
the analysts added 34 new owners to the \db and 11 new abuse types.
They also generalized one abuse type and removed two considered 
to capture techniques rather than abuse types.
The new abuse types appear bolded in Figure~\ref{fig:abuseTags}. 
They include two scam types 
(\ttag{techsupport}, \ttag{phishing}),
malware injecting JavaScript code in a webpage to steal cryptocurrencies (\ttag{web\_skimming}) and
networking indicators being used for malware distribution, vulnerability scanning, ransomware infrastructure, cyberwarfare, data exfiltration, and command-and-control (\ttag{c2}).
A special case of the latter are blockchain addresses used to signal new command-and-control domains or IP addresses (\ttag{blockchain\_signaling}).

Of the \numgtindicators indicators in the GT, 
\numgtiocs (85.4\%) are IoCs (\ttag{abuse}), 
\numgtnotabuse (12.5\%) are benign (\ttag{notabuse}), and 
\numgtnotanioc (2.1\%) are FPs of \iocsearcher (\ttag{notanioc}).
The most common benign indicators are domains and URLs.
Benign domains can be highly popular 
(e.g., \url{avast.io}, \url{bitcoin.org}), 
but also not so popular 
(e.g., \url{ecos.am}, a cryptocurrency mining pool with a Tranco~\cite{tranco} ranking of 428,857).
An example benign URL is 
\url{https://www.torproject.org/download/}.
The most common \iocsearcher FPs are filenames incorrectly identified as 
domain names (e.g., \emph{file.py}, \emph{archive.zip}).
Another common case is filepaths misclassified as URLs,
e.g., \iocsearcher identifies 
\url{transmission.app/Content/MacOS/Transmission} as a URL,
but it is a file inside the \url{transmission.app} mobile app.

Of the 100 reports in the GT, 18 contain IoCs from multiple owners.
These illustrate the need for \tool, as we cannot simply assign all IoCs 
in the report to a single entity.
Instead, these reports need fine-grained, per-IoC processing.
The most common IOC types, abuse types, and owners
are shown in Table~\ref{tab:gt_distribution}.

\begin{table}[t]
\centering
\caption{Top-10 owner, abuse, and IoC types in GT.}
\label{tab:gt_distribution}
\resizebox{\columnwidth}{!}{
\begin{tabular}{rl|rl|rl}
\toprule
\multicolumn{2}{c|}{\textbf{Owner}} & \multicolumn{2}{c|}{\textbf{Abuse Type}} & \multicolumn{2}{c}{\textbf{IoC Type}} \\
\midrule

83 & rtm & 388 & giveaway & 403 & fqdn \\ 52 & xbash & 183 & notabuse & 304 & btc \\
52 & iron-group & 180 & command\_and\_control & 199 & sha256 \\
43 & protonbot & 121 & ransom & 160 & url \\
43 & glad0ff & 106 & ransomware & 121 & ip4 \\
38 & speakup & 68 & clipping & 108 & eth \\
33 & xanthe-elf & 47 & malware\_distribution & 79 & sha1 \\
32 & rocke & 41 & cryptojacker & 62 & md5 \\
30 & combojack & 36 & clipper & 45 & email \\
28 & apt29 & 34 & fake\_marketplace & 35 & doge \\

\bottomrule
\end{tabular}
}
\end{table}

\section{\Tool}
\label{sec:tool}

This section details the main components of \tool, which were introduced in Section~\ref{sec:arch}.

\subsection{Generics Filtering}
\label{sec:generics}

The generics filtering module efficiently identifies a subset of
benign indicators (called \emph{generics}), and tags them as \emph{notabuse}.
This allows skipping the LLM-based tagging modules on those non-IoC indicators,  
improving \tool's efficiency.
The filtering is based on blocklists and does not leverage the report's text. 
Thus, there are benign indicators it cannot identify.
For example, it can identify that popular domains are benign indicators, 
but it cannot determine the same for less popular benign domains. 
It also cannot identify FPs of \iocsearcher such as filenames
(e.g., attachments.zip) 
incorrectly identified as domains.
Benign indicators and \iocsearcher FPs missed by this module will be later identified by the 
abuse tagging module using the report's context.

The determination of whether an indicator is generic is specific 
to each indicator type as described next.

\mypar{Apex domain.}
An apex is generic if it satisfies at least one of these constraints:
(1) appears in the public part of the public suffix list (PSL)~\cite{psl};
(2) appears in the Tranco Top 100K domain popularity list~\cite{tranco};
(3) appears in a public list of 3,791 email providers~\cite{emailProviders}; or
(4) appears in the lists of 6,637 subdomain-leasing and 1,832 URL-leasing apexes 
produced by Ahmed et al.~\cite{ahmed2026blockmenot}.

\mypar{FQDN.}
A fully-qualified domain name (FQDN) is generic if its apex is generic and does not lease subdomains, 
or if its apex leases subdomains but the FQDN does not include a subdomain or the subdomain is www. 

\mypar{URL.}
A URL is generic if it contains a hostname (rather than an IP address) and 
the hostname's apex satisfies one of these constraints: 
(1) it appears in a list of 250 domains owned by security vendors;
(2) it leases URLs and the URL has no path; or 
(3) it is a generic apex.

\mypar{IPv4.}
An IPv4 address is generic if it 
belongs to a private range (e.g., 10.0.0.0/8),
is a multicast address, or 
is in a list of 93 addresses belonging to public DNS resolvers.

\mypar{Email.}
An email is generic if it contains one of 13 generic usernames
(e.g., nobody, noemail)
or if it does not belong to an email provider and the
hostname is a generic FQDN.

Other indicator types 
(i.e., hashes and blockchain addresses)
are considered not generic.

\subsection{Abuse Tagging}
\label{sec:abuse}

The abuse tagging module takes the indicators extracted from the threat report
using \iocsearcher,
filters indicators that are not IoCs, and 
tags the IoCs with abuse type tags.
It addresses four of the five requirements in Section~\ref{sec:problemdef}:
discarding false indicators (R1),
distinguishing benign and malicious indicators (R2), 
refining the IoC type for hashes (R3), and 
assigning abuse type tags (C2).

This module implements an LLM-based classifier that follows the top-to-bottom
design proposed by Gomez et al.~\cite{cleanupthemess}.
While their classifier only supported abuse types for blockchain addresses,
our classifier supports a much larger variety of indicators 
(i.e., contact, hashes, networking, payment),
identifies false indicators, and 
refines the type of indicator a hash captures.

As a preparation step, we wrote a textual definition for each abuse type
in Figure~\ref{fig:abuseTags}.
The definitions are used to force the LLM to interpret the tags 
as we intend to use them.
They prevent the LLM from interpreting the 
tags using its training data.
This is fundamental because abuse types (e.g., scam, extortion)
are often defined differently across sources, 
introducing conflicting definitions that can confuse the LLM.

\begin{figure}
\footnotesize
\centering
\begin{prompt}[Abuse type]

\begin{spverbatim}
You are a cybersecurity expert with extensive knowledge about network and software security, including topics like malware, cryptocurrency scams, advanced persistent threat groups, and other online abuses. You will help me classify Indicators of Compromise (IOCs) that appear in threat reports (i.e., cryptocurrency addresses, IPs, domain names, hashes, etc), based exclusively on the content of the given TEXT.
* Use only information in the TEXT. No external knowledge.
* Do NOT infer beyond explicit statements.
* Do NOT classify based only on similarity, speculation, or historical association.
The following is a LIST OF DEFINITIONS of abuse classes. Read the list carefully and use it to classify the given IOC according to its context within the TEXT, by answering to the given QUESTION.
### LIST OF DEFINITIONS ###
{}

### TEXT ###
{}

### QUESTION ###
Given the LIST OF DEFINITIONS above, classify the IOC {} with IOC type {} using its context (it starts at character {} WITHIN the TEXT) in one of the following classes: {}. Answer only with the name of the class that clearly matches one of these definitions and justify your answer by filling the next JSON structure: {"answer": "", "reasoning": ""}
\end{spverbatim}

\end{prompt}
\caption{Abuse type prompt. Variables in curly brackets are replaced with the tag definitions, the description text, the IoC value, the IoC type, the IoC position in the text, and the list of possible tags to output.}
\label{prompt:abusetype}
\end{figure}

The classifier traverses an abuse type taxonomy,
such as the one in Figure~\ref{fig:abuseTags}, from the top to the bottom. 
At each level, it performs a query to the LLM using the prompt in Figure~\ref{prompt:abusetype}.
The prompt contains variables that capture
the report's text;
the indicator's position, type and value;
the tags that can be output;
and their corresponding definitions.
The difference between queries is that, at each level, 
the classifier includes in the prompt a different
set of tags and definitions.
The possible output tags at the current level depend on the
classification result of the previous level.
For example, if the LLM outputs \ttag{extortion} at the level two (L2) 
classification,
then the definitions for the five extortion children
(e.g., \entry{sextortion}, \entry{reputation}) 
are used in the prompt of the level three (L3) classification.
In addition to the children labels,
an additional \emph{other} label is also included in the prompt.
This label is assigned by the LLM when none of the children labels apply.
The LLM is queried using the prompt, and the process iterates until 
the output is \emph{other}, 
in which case the output label is the label of the previous level, or 
there are no more taxonomy levels to traverse, 
in which case the output label is the label for the last level. 
In addition to the assigned label, 
the output also includes the model's reasoning.

As an optimization on the above process, 
for each top-level indicator type group
(i.e., networking, hashes, contact, blockchain, payment),
we prepare a separate abuse type taxonomy that is a subset of the one in 
Figure~\ref{fig:abuseTags}, 
where tags that do not apply to that group have been removed. 
For example, \ttag{malware\_distribution} and \ttag{vulnerability\_scanning}
are only included in the taxonomies for networking indicators. 
This limits the options that the LLM needs to reason about, 
improving accuracy, and reduces the size of the queries, 
reducing the classification cost.
The used taxonomies are detailed in Figures~\ref{fig:metataxonomy_networking}--\ref{fig:metataxonomy_blockchain} in the Appendix.

\subsection{Owner Tagging}
\label{sec:owner}

The owner tagging module produces a set of owner tags
that capture the entities controlling the IoCs,
satisfying requirement C1.
Not every IoC is assigned an owner. 
IoCs whose abuse type is \ttag{extortion} or \ttag{scam}, 
or any of their children in Figure~\ref{fig:tags}, are not assigned an owner 
because those types of abuse are not directly associated with 
malware families or threat groups. 
The exception is investment scams,
for which the owner may be
a malicious company or individual that offers the investment,
rather than a malware family or threat group.

Owner tagging comprises two steps.
First, the LLM is queried to \emph{propose} a set of owner names 
in an open-world fashion,
i.e., without constraining the names the model may output.
Second, a \emph{refinement} step tries to identify for each 
proposed name a corresponding entity in the \db,
taking into account known aliases.
If an entity is found, its canonical name is used as a tag. 
Otherwise, a new tag is created. 
We detail both steps next.

\mypar{Owner identification.}
For identifying the owner,
\tool uses a \emph{chain of thought} prompt.
The prompt first defines what we mean by ownership.
Then, it includes a set of rules that must be followed, 
which aim at minimizing speculation and hallucinations
by forcing the model to reason about the text
instead of solely relying on its training knowledge.
The rules also detail some entities that should be ignored because, 
while they may be considered the owners of an IoC, 
in reality they do not control how the IoC is used because they have 
leased it to some other entity in exchange for a fee, 
e.g., hosting providers and domain registrars.
Next, the prompt explains that the task should be solved by performing 
two actions. 
First, the model should identify the type of the entity 
that controls the indicator among the following five types:
\ttag{malware\_family}, \ttag{threat\_group}, \ttag{individual}, 
\ttag{organization}, and \ttag{government}.
Then, it requests the model to identify the name of the owner given its type, 
explaining that, if no owner can be identified, it should return 
\ttag{unknown}, and if multiple, equally likely, owners exist, 
it should return a list with all of them.
The prompt also constrains the format of the provided name to use only 
lowercase characters, digits, hyphens, underscores, dots, and spaces.
Finally, it asks for the resulting owner(s) to be output in JSON format
and to include the reasoning.
In the example in Figure~\ref{fig:running_example},
\tool proposes two candidate owner labels
for the domain name \ttag{com-ae.net}:
the malware family \ttag{prospy} and the threat group \ttag{bitter apt}.

\mypar{Owner refinement.}
The goal of owner refinement is to check if the names proposed by the 
LLM may correspond to entities already present in the \db.
For this, it needs to check whether the LLM may have proposed an alias for 
an existing entity.
If so, the tag (i.e., canonical name) of the existing entity should be reused, 
avoiding the creation of multiple tags for the same entity.

Owner refinement first uses the owner type obtained in owner 
identification to split the proposed names into three:
threat groups, malware families, and other owners, 
with the latter comprising individual, organization, and government types.
Then, for each of these three groups, it iterates on the names 
proposed by the LLM searching if the name already exists in the \db
either as a canonical name (using the search\_entity function) 
or as an alias (using the search\_alias function).
In our example, when processing \url{com-ae.net}, 
neither \ttag{prospy} nor \ttag{bitter apt} are canonical names or known 
aliases in the \db.
If those two searches fail, 
it uses the search\_k\_similary function to try to identify similar 
canonical names or aliases in the \db. 
As explained in Section~\ref{sec:entities}, this function returns 
at most $k$, but possibly fewer, candidate entities.
If it returns no candidate entities, 
the name proposed by the LLM is output as a new tag. 
If it returns a single candidate, then the canonical name of that candidate 
is used as a tag. 
Finally, if multiple candidates are output by the similarity search, 
\tool selects the best among the proposed name and these candidates
by querying the LLM with a refinement prompt.
In our example, the similarity search finds no candidates for \ttag{prospy}, 
so a new entity is created with this name as the canonical name and used as a tag.
For the \emph{bitter apt} proposed name, the similarity search finds 
two candidates: the \ttag{bitter} threat group and the \ttag{bitter-rat}
malware family.
The LLM is invoked to decide which of these two candidates best matches
the proposed name.

The refinement prompt includes
the IoC formatted with its IoC type, 
the label proposed by the LLM during the owner identification step
and the reasoning for selecting it, and
the candidate entities. 
By including the reasoning provided by the LLM
in the owner identification step,
we force the LLM to reason about the candidates, 
instead of simply choosing the candidate with highest syntactic similarity.
The LLM may choose the best-fit candidate, 
in which case the canonical name of that candidate is used as a tag.
Alternatively,  it may decide that none of the candidates is a good match. 
In that case, a new entity is created with its canonical name being the 
LLM's proposed name, and the proposed name is used as tag.
In our example, the LLM chooses \ttag{bitter} as the best match for 
\ttag{bitter apt}, reusing an existing tag.

\section{Evaluation}
\label{sec:evaluation}

This section evaluates \tool. 
Section~\ref{sec:accuracy} first presents accuracy results on the GT. 
Then, Section~\ref{sec:measurement} presents results of 
applying \tool to tag new reports.

\subsection{Accuracy Evaluation}
\label{sec:accuracy}

We first evaluate \tool's accuracy for producing abuse type and owner tags
using the GT built in Section~\ref{sec:gt}.
We have tested \tool with four open model families: 
OpenAI's gpt-oss (20b),
Google's gemma4-e (4b),
DeepSeek R1 (5b, 7b, 32b), and 
Qwen 3.5 (4b, 9b, 27b). 
However, the DeepSeek R1 and Qwen 3.5 models frequently failed to output a 
properly formatted JSON, 
or ran too slowly on our Debian server,
equipped with an NVIDIA A100 GPU with 40GB of vRAM,
256 GB of RAM,
and 64 CPUs.
Thus, we present results only for two models:
OpenAI's gpt-oss-20b and Google's gemma4-e4b.

We evaluate our tagging modules as multi-class classifiers.
We first compute precision, recall, and F1 score for each class (one-vs-all).
Then, we calculate the weighted average of each metric,
where each class is weighted by the number of samples in the class.
We use a strict view of accuracy which considers a true positive (TP)
only if \tool outputs exactly one of the tags for the IoC in the GT.
For example, if the GT says \ttag{sextortion} and \tool
outputs the parent \ttag{extortion}, we consider it an error.

\begin{table}[t]
\centering
\caption{Abuse type tagging accuracy on GT for each taxonomy level 
and final accuracy when using all four levels.}
\label{tab:accuracy_abuse_type}
\begin{tabular}{l|c|ccc}
\toprule
Model & Level & Precision & Recall & F1 \\
\midrule

gpt-oss-20b & L1 & 0.99 & 0.99 & 0.99 \\
gemma4-e4b & L1 & 0.97 & 0.97 & 0.97 \\
\midrule
gpt-oss-20b & L2 & 0.96 & 0.95 & 0.95 \\
gemma4-e4b & L2 & 0.91 & 0.84 & 0.87 \\
\midrule
gpt-oss-20b & L3 & 0.96 & 0.94 & 0.95 \\
gemma4-e4b & L3 & 0.90 & 0.77 & 0.81 \\
\midrule
\midrule
gpt-oss-20b & All & 0.94 & 0.93 & 0.93 \\
gemma4-e4b & All & 0.87 & 0.72 & 0.76 \\

\bottomrule
\end{tabular}
\end{table}

\mypar{Abuse type tagging.}
Table~\ref{tab:accuracy_abuse_type} summarizes \tool's abuse tagging accuracy, 
as well as the accuracy for each of the four taxonomy levels.
The best \tool accuracy is achieved when using the gpt-oss-20b model.
Using that model, \tool achieves an abuse tagging 
precision of 0.94, a recall of 0.93, and an F1 score of 0.93.
In contrast, when using gemma4-e4b, all metrics are lower, 
achieving a precision of 0.87, recall of 0.72, and F1 score of 0.76. 

For the per-level results, we allow \tool to output just the abuse types 
at that level.
For instance, at L1, \tool only assigns the 
\ttag{abuse}, \ttag{notabuse}, and \ttag{notanioc} tags.
The results show that gpt-oss-20b outperforms gemma4-e4b at all levels, and
that, regardless of the model, 
the accuracy reduces as the classification becomes more fine-grained.
For example, when using gpt-oss-20b, the F1 score is 
0.99 at L1, 0.95 at L2, 0.95 at L3, and 
finally 0.93 when all four levels are used.
This happens because errors accumulate, 
i.e., an error in upper levels cannot be corrected.

Figure~\ref{fig:gt_confusion_matrix} in the Appendix shows the confusion 
matrix for abuse type tags when using gpt-oss-20b.
We observe three common sources of errors.
The first case is \ttag{giveway} 
blockchain IoCs misclassified as \ttag{unauthwithdrawal} (20 cases) and 
\ttag{phishing} networking IoCs misclassified as \ttag{giveaway} (13 cases). 
A single report~\cite{arkinvest} 
contains 10 misclassified Bitcoin addresses where 
the LLM's reasoning states that the Bitcoin addresses were used to 
receive \emph{stolen} funds from a giveaway scheme,
even if the reasoning also states that 
``the victims \emph{voluntarily} sent the funds to double them''. 
The LLM fails to reason that the funds are not withdrawn without authorization
and incorrectly concludes that ``this is an unauthorized withdrawal of funds''.
Still, both types of errors affect less than 9\% of 
the 367 \ttag{giveaway} IoCs.
The second common cases are \ttag{malware\_distribution} networking IoCs 
identified as \ttag{command\_and\_control} (5 cases) and the opposite case (4).
We meant \ttag{malware\_distribution} to capture only the distribution
of the first malware sample (the one that infects the victim's host) 
and any subsequent malware downloads to be tagged 
as coming from \ttag{command\_and\_control} servers. 
But, this distinction seems to be too subtle. 
An alternative could be to move \ttag{malware\_distribution} as a child of 
\ttag{command\_and\_control}.
The third case consists of discrepancies between the malware categories assigned 
underneath \ttag{malware\_sample}. 
In some of these, the analyst assigned
the same abuse type of the malware family (e.g., \ttag{ransomware}),
to different malicious files belonging to the same malware family,
but implementing different components
(e.g., \ttag{downloader}, \ttag{infostealer}).
However, the LLM tries to make the classification more fine-grained, 
e.g., stating that some of the ransomware components are downloaders.

\begin{table}[t]
\centering
\caption{Owner tagging accuracy on GT.}
\label{tab:accuracy_owner}
\begin{tabular}{lccc}
\toprule
Model & Precision & Recall & F1 \\
\midrule
gpt-oss-20b & 0.95 & 0.94 & 0.94 \\
gemma4-e4b	& 0.79 & 0.76 & 0.73 \\
\bottomrule
\end{tabular}
\end{table}

\mypar{Owner tagging.}
Table~\ref{tab:accuracy_owner} summarizes the accuracy of \tool when 
generating owner tags.
Since owner tagging is skipped for IoCs of certain abuse types 
(i.e., extortions and scams), 
we consider that the assigned owner tag is correct only if the abuse type 
was correct and the owner tags assigned by \tool have a non-empty intersection 
with those in the GT. 
Again, the best accuracy is achieved using gpt-oss-20b, 
which achieves a precision of 0.95, a recall of 0.94, and an F1 score of 0.94,
compared to a precision of 0.79, recall of 0.76, and an F1 score of 0.73 
when using gemma4-e4b.

Of the 838 IoCs that are undergoing owner tagging, 
784 (93.6\%) are assigned an owner. 
For the other 54 (6.4\%), the LLM returns ``unknown''.
Most FNs come from reports where
the LLM only extracted an owner for a subset of IoCs, 
despite multiple of them potentially sharing the same context
(i.e., being part of the same table), 
an inconsistency problem likely due to the non-deterministic nature of LLMs.

Among the FPs, the most common case is file hashes appearing in the 
report with the label assigned to the sample by an AV engine, 
e.g., Win32/TrojanDownloader.Banload.YJD.
The LLM tends to output this label as the malware family, 
while the analyst tends to add to the GT a more specific family name 
(e.g., casbaneiro).
While AV labels may contain a family name, 
they often contain other information unrelated to the family
(e.g., Win32, TrojanDownloader, YJD)
which malware labeling tools can filter~\cite{avclass}.
Integrating such tools into \tool is a possible avenue for future work.

\begin{table}[t]
\centering
\caption{Ablation results comparing the accuracy of \tool using gpt-oss-20b 
with versions that remove the generics identification and the 
owner normalization steps.
}
\label{tab:ablation}
\begin{tabular}{l|ccc|ccc}
\toprule
\multicolumn{1}{l|}{} & \multicolumn{3}{c|}{\textbf{Abuse Type}} & \multicolumn{3}{c}{\textbf{Owner}} \\
\midrule
System & Prec & Rec & F1 & Prec & Rec & F1 \\
\midrule
\tool & 0.94 & 0.93 & 0.93 & 0.94 & 0.93 & 0.93 \\
No generics  & 0.94 & 0.92 & 0.92 & 0.94 & 0.93 & 0.93 \\  
No owner normalization & 0.94 & 0.93 & 0.93 & 0.85 & 0.81 & 0.82 \\  
\bottomrule
\end{tabular}
\end{table}

\mypar{Ablation study.}
We evaluate the impact of removing different components of \tool, 
in particular the generics identification and owner normalization modules.
Table~\ref{tab:ablation} summarizes the ablation study results.
Removing the generics identification module 
reduces the abuse tagging F1 score from 0.93 to 0.92. 
This version of \tool uses the LLM to classify every indicator 
output by \iocsearcher, making the whole pipeline run slower, 
as it queries the LLM for more indicators 
(i.e., the generics module filters 9.6\% of all GT indicators), but also being less accurate.
This happens due to indicators correctly identified as benign 
by the generics identification module
being incorrectly classified as IoCs by the LLM.
For example, the domain download.cnet.com is correctly identified as 
benign by the generics module. 
Instead, when not using the generics module, the LLM reasons that 
``download.cnet.com hosted trojanized applications that steal bitcoin; the domain is used as a distribution point for malicious software, which is a clear instance of abuse'', making the domain be incorrectly considered an IoC.

When removing the owner normalization module,
the owner tagging F1 score decreases from 0.93 to 0.82.
This happens because the raw owner label output by the LLM may be an alias 
such as \emph{msil\_clipbanker\_df} instead of \ttag{clipbanker} or
\ttag{redaman} instead of \ttag{rtm}.

\begin{table}[t]
\centering
\caption{Level 1 (\ttag{abuse}, \ttag{notabuse}) classification results 
on the PRISM dataset~\cite{froudakis2025revealing} using 
only the generics module, only gpt-oss-20b, and both components.
\ttag{notanioc} is considered part of \ttag{notabuse} for this experiment.
}
\label{tab:prism}
\begin{tabular}{l|cccr}
\toprule
\textbf{L1 Classifier} & \textbf{Precision} & \textbf{Recall} & \textbf{F1} & \textbf{LLM Queries} \\
\midrule
Generics only & 0.98 & 0.55 & 0.70 & - \\

LLM L1 only & 0.98 & 0.98 & 0.98  & 1,475 \\ Generics + LLM L1 & 0.98 & 0.98 & 0.98 & 1,274 \\ 
\bottomrule
\end{tabular}
\end{table}

\mypar{L1 classification accuracy.}
We also evaluate the accuracy of the top-level (L1) classification, 
which is fundamental for the overall accuracy since errors accumulate. 
Froudakis et al.\cite{froudakis2025revealing} recently released the 
PRISM dataset with 1,746 indicators extracted by \iocsearcher from 
50 threat reports, 
with each indicator labeled as an IoC 
or not.
Table~\ref{tab:prism} shows the accuracy of the L1 classification on the PRISM dataset when using: 
only the generics module, 
only the L1 prompt with gpt-oss-20b, 
and both components.
Since PRISM does not differentiate \ttag{notanioc}, we consider that tag as \ttag{notabuse}
for this experiment.

The generics identification module achieves a 
precision of 0.98, a recall of 0.55, and an F1 score of 0.70.
The very high precision (0.98) validates the use of the 
generics module by \tool to reduce the number of queries to the LLM.
There are only four indicators labeled as potential FPs. 
Manual examination of those reveals that two are indeed FPs of our module
due to apexes that lease URLs only under certain paths
(i.e., \url{https://discord.com/api/webhooks/<WEBHOOK>} or subdomains
(i.e., \url{https://graph.microsoft.com/beta/users/<UUID>}).
The other are not really FPs but errors in the PRISM GT corresponding to 
a domain (a.storyblok.com) and a URL (\url{https://notifier.rarlab.com}) 
flagged as IoCs but belonging to benign services.
On the other hand, the recall of the generics module is low 
due to indicators that can only be determined not to be an IoC by 
analyzing the report's text. 
Common cases are filenames 
(e.g., attachments.zip, framework.py) 
and Android package names 
(e.g., com.avira.android, com.bitdefender.security) 
incorrectly identified as domains by \iocsearcher.

Using only the LLM already achieves a precision of 0.98, recall of 0.98, and F1 score of 0.98.
However, it requires doing one LLM query for each candidate indicator. 
By using the generics module, we save 12\% of LLM queries on PRISM, 
but, as we will show in Section~\ref{sec:measurement}, 
in other datasets like AnnoCTR more than half the candidate indicators are benign, 
producing much larger savings in LLM queries.
Furthermore, when using the full \tool, each benign indicator identified by the generics modules
can save multiple queries to the LLM.  
When combining both the generics and LLM L1 components,
\tool achieves an L1 precision of 0.98, recall of 0.98, and F1 score of 0.98, 
while saving a significant number of LLM queries.

Among the 152 potential FNs of the generics identification module, 
there are 26 domain names with at least four detections in VirusTotal 
(e.g., abert-online.de, acehigh.host, androidd.com).
These domains are highly likely malicious, 
but incorrectly marked as not IoCs in PRISM.
An additional 6 domains have 2--3 VT detections; 
these could be malicious, but also FPs from AV engines.
We will report these cases to the PRISM authors.

\begin{table*}[t]
\centering
\caption{Summary of results when applyting \tool to Malpedia, AnnoCTR, and PRISM reports.}
\label{tab:measurement}
\resizebox{\textwidth}{!}{
\begin{tabular}{l|rrr|rrrr|r|rrr}
\toprule
& \multicolumn{3}{c|}{\textbf{Reports}} & \multicolumn{4}{c|}{\textbf{Indicators}} & \multicolumn{1}{c|}{\textbf{IoCs}} & \multicolumn{3}{c}{\textbf{Owner Entities}} \\
\textbf{Dataset} & \textbf{All} & \textbf{w/Ind.} & \textbf{w/IoCs} & \textbf{All} & \textbf{notanioc} & \textbf{notabuse} & \textbf{IoCs} & \multicolumn{1}{c|}{\textbf{w/Owner}} & \textbf{All} & \textbf{Known} & \textbf{New} \\
\midrule

Annoctr & 400 & 386 (96.5\%) & 273 (68.2\%) & 8,804 & 123 (1.4\%) & 2,933 (33.3\%) & 5,749 (65.3\%) & 4,118 (71.6\%) & 364 & 210 & 154 \\
Malpedia & 315 & 272 (86.3\%) & 209 (66.3\%) & 9,912 & 379 (3.8\%) & 1,076 (10.9\%) & 8,476 (85.5\%) & 5,695 (67.2\%) & 461 & 190 & 271 \\
PRISM & 50 & 50 (100.0\%) & 50 (100.0\%) & 1,590 & 65 (4.1\%) & 167 (10.5\%) & 1,359 (85.5\%) & 1,149 (84.5\%) & 131 & 83 & 48 \\

\midrule
All & 765 & 708 (92.5\%) & 532 (69.5\%) & 20,239 & 563 (2.8\%) & 4,117 (20.3\%) & 15,583 (77.0\%) & 10,961 (70.3\%) & 912 & 439 & 473 \\
\bottomrule
\end{tabular}
}
\end{table*}

\subsection{Report Tagging}
\label{sec:measurement}

In this section, we apply \tool to tag 765 threat reports from three sources:
400 reports from the AnnoCTR dataset~\cite{annoctr},
published between March 2013 and February 2022;
50 reports from the PRISM dataset~\cite{lance},
published between April 2023 and November 2024; and
315 reports added to Malpedia~\cite{malpedia}
between February 18 and May 28, 2026.
Of the Malpedia reports, 70\% were published in 2026, 21\% in 2025, and 9\% are older reports,
going as far back as 2016.
According to the report URLs and titles, the reports from the three sources do not overlap.

Table~\ref{tab:measurement} summarizes the tagging results.
Of the 765 reports analyzed, 708 (92.5\%) contain at least one candidate
indicator identified by \iocsearcher.
The rest are mostly Malpedia reports that describe techniques used by malware but 
do not reference specific IoCs (e.g.,~\cite{cocomelonc, linuxrootkits}).
Of the 20,239 candidate indicators,
\tool tags 15,583 (77.0\%) as IoCs. 
4,117 (20.3\%) as benign, and 
563 (2.8\%) as FPs of \iocsearcher.
The IoCs come from 532 (69.5\%) of the analyzed reports. 
Reports without IoCs include quarterly reports where security vendors
compare the prevalence of different threats,
without providing IoCs for those threats.
The ratio of benign indicators in AnnoCTR (33.3\%) is significantly higher
than in Malpedia (10.9\%) and PRISM (10.5\%).
This happens due to AnnoCTR including URL references 
(e.g., citations to other threat reports) as part of the report's text,
while in the other two datasets we extracted the text directly from the HTML and PDF 
reports using \iocsearcher,
which includes the text of links, but not the link URLs, as part of the text.

\begin{table}[t]
\centering
\caption{Top 10 abuse type tags by number of IoCs.}
\label{tab:top_abuse}
\begin{tabular}{lrrrr}
\toprule
\textbf{Tag} & \textbf{IoCs} & \textbf{Malp.} & \textbf{Anno.} & \textbf{PRISM} \\
\midrule

command\_and\_control & 4,191 & 2,069 & 1,706 & 417 \\
malware\_distribution & 3,422 & 1,806 & 1,460 & 156 \\
phishing & 1,959 & 1,442 & 452 & 65 \\
downloader & 1,192 & 512 & 492 & 188 \\
infostealer & 1,022 & 191 & 756 & 75 \\
rat & 891 & 439 & 214 & 238 \\
registration\_contact & 702 & 698 & 3 & 1 \\
abuse & 656 & 545 & 97 & 14 \\
malware\_sample & 632 & 234 & 258 & 140 \\
sender & 372 & 258 & 111 & 3 \\
file\_hash & 184 & 71 & 61 & 52 \\

\bottomrule
\end{tabular}
\end{table}

\mypar{Abuse type tags.}
\Tool assigns an abuse type tag to each of the 15,583 IoCs.
Of those, 95.8\% are assigned a specific abuse type,
while the other 4.2\% are assigned the top-level \ttag{abuse} tag,   
indicating that no specific abuse type could be found. 
Most of these IoCs (83.1\%) come from Malpedia.
The large majority are
domain names and IP addresses
used in misinformation campaigns
(e.g., to host fake news,
launch political influence campaigns, or
spread propaganda),
or that are part of unspecified malicious infrastructure.
This points towards \ttag{cyberwarfare} being a tag that should also be 
applied to networking indicators.

Table~\ref{tab:top_abuse} shows the top 10 abuse types. 
The most common are 
\ttag{command\_and\_control} assigned to 26.9\% IoCs,
\ttag{malware\_distribution} (22.0\%), and 
\ttag{phishing} (12.6\%), all of which are assigned to networking IoCs.
Also common are malware classes assigned to samples such as
\ttag{downloader} (7.6\%),
\ttag{infostealer} (6.6\%), and
\ttag{rat} (5.7\%).
There are also malicious executables where no specific malware class could be assigned 
(\ttag{malware\_sample}) and benign tools and non-executable files (\ttag{file\_hash}).
An interesting case is sender addresses in malicious emails (\ttag{sender\_email}). 
Since sender addresses are oftentimes spoofed, this tag allows filtering such indicators.

\mypar{Owner tags.}
Among the \numiocs IoCs, \tool assigns an owner tag to \numiocsowner (70.3\%). 
IoCs not assigned an owner belong to abuse types where ownership 
is not considered, such as extortions, phishing, and other scams.
The \numiocsowner IoCs with an owner belong to \numentities entities.
Of those, 439 (48.1\%) were in the original \db 
and the other 473 (51.9\%) are new entities, not in the \db.
Of all entities,
\numgroups are threat groups, and \numfamilies are malware families.
The other \numotherowners entities belong to
individuals,
accounts used as proxies for individuals,
states (e.g., Russia, DPRK), or 
national agencies (e.g., FBI, SVR). 
Of the 912 entities, 
8 appear in reports from the three sources
(e.g., \ttag{apt28}, \ttag{lazarus}, \ttag{agent-tesla}) 
and another 26 appear in reports from two datasets.

\begin{table}[t]
\centering
\caption{Top 10 owner entities by number of IoCs.
}
\label{tab:top-entities}
\begin{tabular}{llrrc}
\toprule
\textbf{Entity} & \textbf{Type} & \textbf{IoCs} & \textbf{Reports} & \textbf{Known} \\
\midrule

hazy-tiger & Group & 387 & 1 & \Y \\
apt28 & Group & 369 & 14 & \Y \\
north korean nation state & Group & 354 & 1 & \N \\
scripted-sparrow & Group & 332 & 1 & \Y \\
graycharlie & Group & 315 & 1 & \Y \\
prospy & Malware & 285 & 1 & \N \\
storm-1516 & Group & 204 & 1 & \Y \\
unc2814 & Group & 197 & 1 & \Y \\
emotet & Malware & 189 & 8 & \Y \\
flubot-apk & Malware & 177 & 1 & \Y \\

\bottomrule
\end{tabular}
\end{table}

Table~\ref{tab:top-entities} 
shows the top-10 owner entities by number of IoCs identified 
across reports from all sources. 
Among these 10 entities, 7 are threat groups, and 3 are malware families.
Some entities like \ttag{hazy-tiger}, \ttag{scripted-sparrow}, or \ttag{graycharlie}
have a single report with many IoCs,
but others have IoCs coming from multiple reports such as 
\ttag{apt28} (14 reports from all three sources) and \ttag{emotet} (8 reports from Malpedia).  
Eight of these entities were known.
The unknown entities come from two reports,
one of malicious activities linked to North Korea
where the authors refer to the malicious actor as
\ttag{North Korean nation-state threat actors}.
The other comes from the report in our example in Figure~\ref{fig:running_example}
that mentions a previously unknown malware named \ttag{prospy}.

\section{Discussion}
\label{sec:discussion}

We discuss limitations and avenues for improvement. 

\mypar{Indicator extraction.}
We use \iocsearcher to identify candidate indicators in threat reports and 
then apply \tool to tag those. 
Thus, any indicator that \iocsearcher misses introduces an FN. 
While \iocsearcher has been shown to have very high 
recall~\cite{goodfatr,froudakis2025revealing}, 
it occasionally misses indicators 
such as when URLs are split between consecutive lines.
We have preliminarily tested extracting indicators from the threat reports 
directly with the LLM, without using \iocsearcher. 
But the recall was lower because the LLM refused to be thorough. 
Still, we believe direct indicator extraction with LLMs, 
without regular expressions, is a promising approach.

\mypar{Multiple abuse types.}
\Tool may extract multiple owner tags for an IoC in a report, 
but currently only extracts a single abuse type tag per IoC. 
An IoC may be used for different types of abuse. 
When the same IoC is mentioned in different reports, we can accumulate tags 
for the IoC.
Extracting multiple abuse type tags from the same report may require 
changing our hierarchical taxonomy approach, which is critical to making 
the taxonomy support new abuse types.
This is an important area for future work.

\mypar{Open models.}
\Tool is independent of the LLM used. 
We have evaluated it using a variety of open models run locally.
Using commercial frontier models may further improve its accuracy,
but would make our evaluation less replicable and more costly.

\mypar{Images.}
Currently \tool only analyzes the report's text. 
In some cases, IoCs and their context may only appear in images.
In future work, we plan to explore multimodal LLMs that can be provided 
with the full report, e.g., the PDF file with both text and images. 

\mypar{Human in the loop.}
We have designed \tool to be fully automated. 
However, humans may want to supervise its correct operation. 
There are two steps where we envision that human supervision may help. 
First, a human could check the new tags that \tool proposes to create,
i.e., when owner refinement does not find any candidate in the \db.
The human may want to check that the new tags are specific enough, 
filtering wide tags such as \emph{north korean nation state} 
in Table~\ref{tab:top-entities}.
Second, \tool may occasionally output multiple owner types for the same 
owner type, e.g., two malware families for the same IoC. 
This may happen when the report's text is confusing. 
\Tool can automatically mark such cases so that a human can double-check them.

\section{Related Work}
\label{sec:related}

Cyber Threat Intelligence (CTI) extraction has evolved
from rule and regular-expression-based IoC extraction
toward contextual information extraction, relation extraction, and,
most recently, large language model (LLM)-based approaches
that aim to transform unstructured threat reports
into structured, actionable intelligence,
with prior work focusing primarily on extracting
indicators of compromise (IoCs),
named entities,
attack behaviors, and
ATT\&CK tactics and techniques.

Early systems
~\cite{liao2016acing,ttpdrill,malpedia,chainsmith,iocminer,timiner,extractor}
focus on extracting structured indicators from unstructured text
by using regular expressions,
which capture the structure of the indicators
(e.g., hashes, IP addresses).
In~\cite{goodfatr},
authors present \textit{iocsearcher}~\cite{iocsearcher},
an indicator extraction tool that identifies 41 different IoC types,
and GoodFATR~\cite{goodfatr},
a threat report collection platform
that evaluates different indicator extraction tools against each other
with no need for a ground truth.
Although authors show that iocsearcher outperforms similar tools when extracting indicators,
it is very limited when it comes to deciding if an indicator is malicious (IoC) or not (generic).
\tool goes beyond iocsearcher's rule-based filtering
and leverages the context of the indicators within the report
to filter out benign indicators
and iocsearcher's false positives
(e.g., a file name artifact extracted as a domain name).

Some approaches aim to extract additional useful information from the reports,
like threat actions~\cite{ttpdrill}, attack behaviors~\cite{extractor},
attack techniques~\cite{attacktech}, and,
more recently, MITRE ATT\&CK tactics and techniques~\cite{attackCTI,ttpxhunter},
extracting semantic meaning from the context of the indicators
by combining NLP with techniques like deep learning (e.g., BERT).
Other approaches transform extracted information into structured representations,
like knowledge graphs of attack techniques~\cite{attackg},
graphs enriched with temporal patterns of attacks~\cite{miningCTI}, and
standardized STIX objects~\cite{stix}.
More recently,
LLM-based methods have been applied to both extraction and classification of indicators
using different prompting techniques~\cite{CTIcontextual,CTIEchoChamber,cleanupthemess}.
In~\cite{cleanupthemess}, authors leverage LLMs to
identify the context of indicators of compromise (i.e., abuse reports)
to extract their abuse type,
effectively separating malicious indicators from benign ones.
However, authors focus on classifying abuse reports
instead of indicators, and do not try to extract any ownership information,
as \tool does.

Research has also examined the broader threat-reporting ecosystem,
including fragmentation and information overlap across long-term CTI reporting~\cite{CTIEchoChamber},
identifying ontology ambiguity and dataset quality as major challenges~\cite{autoCTI}.
Annotated resources such as AnnoCTR~\cite{annoctr} and PRISM~\cite{lance} have been introduced
to support research of entities, tactics, and techniques in CTI reports.
Despite this extensive body of work,
existing approaches predominantly focus on attack-technique extraction or entity recognition
rather than understanding the context and role of individual indicators.
In particular, they generally do not jointly determine an IoC's fine-grained
abuse type and the entity controlling or owning it,
and many implicitly assume that a report concerns a single threat actor or campaign.
In contrast,
\tool addresses the contextual classification of individual IoCs and their controlling entities,
including reports involving multiple threat actors,
by jointly performing fine-grained IoC abuse-type classification
and entity ownership attribution at scale.

\section{Conclusion}
\label{sec:conclusion}

This paper has presented \tool,
a platform for automatically annotating
IoCs in threat reports
with owner and abuse type tags.
Unlike prior work that
focuses primarily on extracting attack techniques and named entities,
\tool focuses on identifying per-IoC owner entities and abuse types,
enriching IoCs with actionable threat intelligence
that can answer interesting questions,
for instance, who owns an IoC
and for what type of cyberattacks the IoC is used.
\Tool can identify IoC abuse types and owner entities
even in reports that describe multiple entities.

Our evaluation demonstrates that
\tool accurately generates owner and abuse type tags,
achieving an F1 score of 0.94 and 0.93, respectively.
We applied \tool to tag \numwildreports threat reports,
identifying \numiocs IoCs belonging to
\numfamilies malware families,
\numgroups threat groups, and
\numotherowners other entities.
\Tool enables the generation of threat group and malware family IoC 
profiles that can be used for threat attribution, threat hunting,
infrastructure tracking, and automated correlation across campaigns.

{
	\bibliographystyle{abbrv}
	% \balance
	\bibliography{bibliography/paper}
}

\appendix

\section{Abuse Type Taxonomies}
\label{sec:taxonomies}

\begin{figure}[t]
\centering
\includegraphics[scale=0.3]{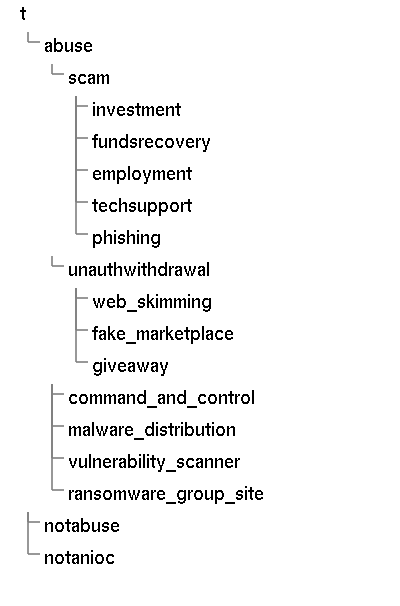}
\caption{Networking taxonomy.}
\label{fig:metataxonomy_networking}
\end{figure}

\begin{figure}[t]
\centering
\includegraphics[scale=0.3]{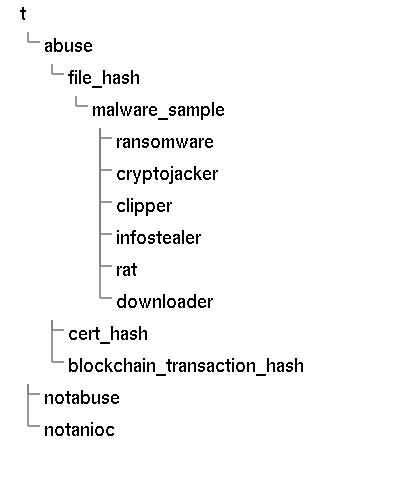}
\caption{Hashes taxonomy.}
\label{fig:metataxonomy_hashes}
\end{figure}

\begin{figure}[t]
\centering
\includegraphics[scale=0.3]{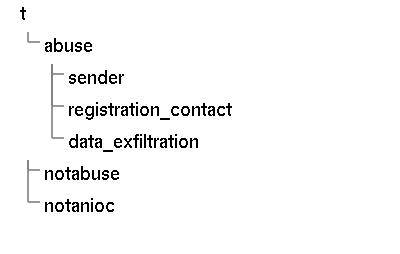}
\caption{Contact taxonomy.}
\label{fig:metataxonomy_email}
\end{figure}

\begin{figure}[t]
\centering
\includegraphics[scale=0.3]{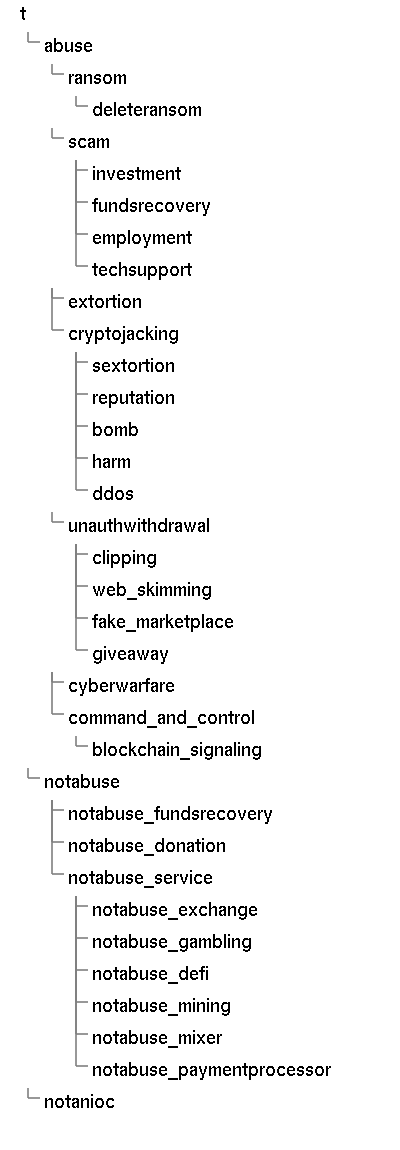}
\caption{Blockchain taxonomy.}
\label{fig:metataxonomy_blockchain}
\end{figure}

The taxonomies used by \tool are detailed in Figures
~\ref{fig:metataxonomy_networking},
~\ref{fig:metataxonomy_hashes},
~\ref{fig:metataxonomy_email}, and
~\ref{fig:metataxonomy_blockchain}.

\section{Category Evaluation}
\label{sec:cateval}

\begin{figure*}[t]
\centering
\includegraphics[width=\textwidth]{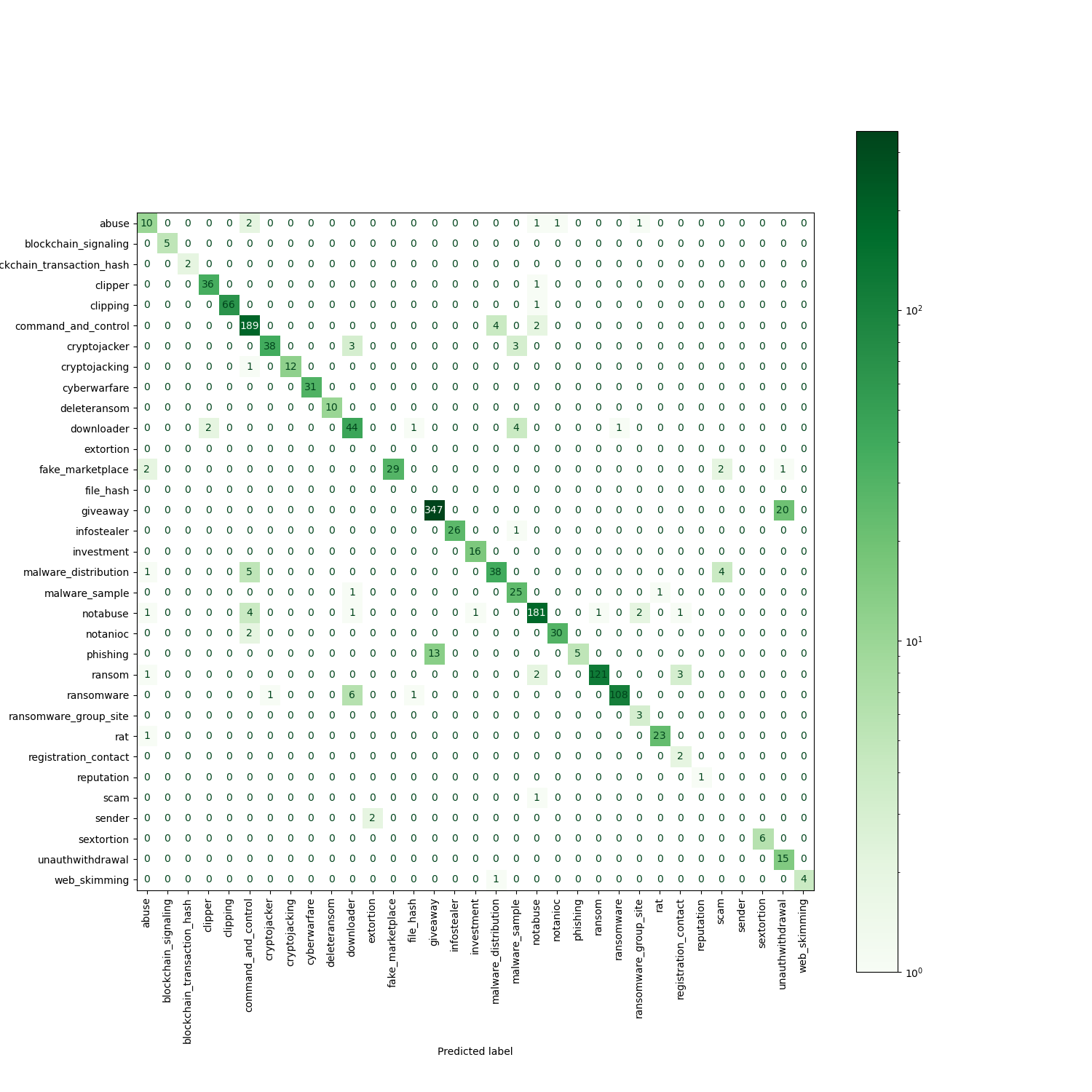}
\caption{Confusion matrix for abuse type in the GT.}
\label{fig:gt_confusion_matrix}
\end{figure*}

Figure~\ref{fig:gt_confusion_matrix} shows the confussion matrix for the abuse type
when classifying the GT using \tool.

\clearpage

\end{document}